\documentclass[aps,prd,twocolumn,showpacs,eqsecnum,nofootinbib]{revtex4}

\usepackage{graphicx}
\usepackage{amsmath}

\font\FermiPPTfont=cmssbx10 scaled 1440
\font\FermiSmallfont=cmssq8 scaled 1200

\def\FNALppthead#1#2#3{
\null 
\begin{center}\vskip -1.0truein{\hbox to 7.5truein {
\vbox to 1in{\vfill 
             \hbox{\includegraphics[height=1.5cm]{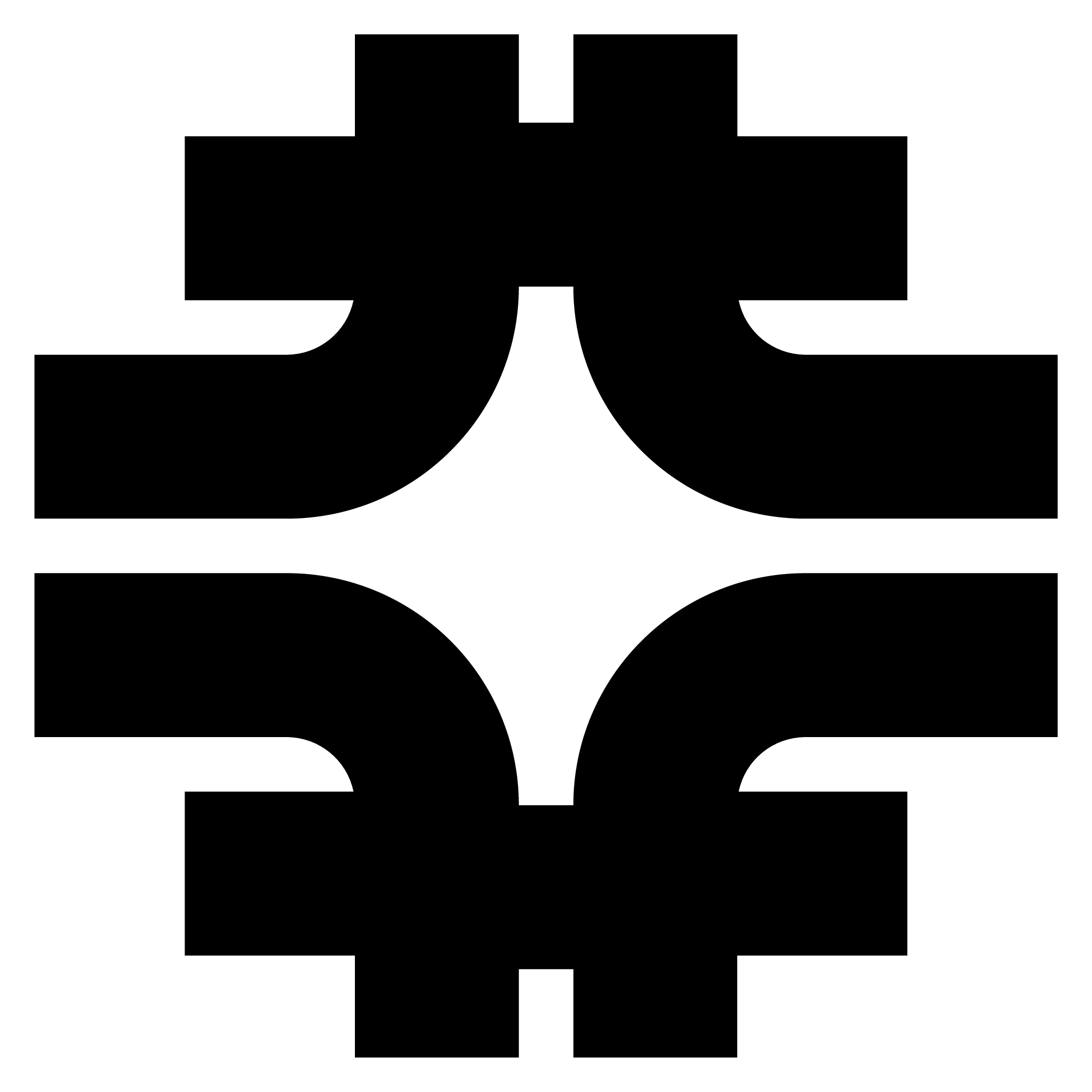}}
             \vfill }
\hskip 1em
\vbox to 1in{\vfill
             \hbox{{\FermiPPTfont Fermi National Accelerator Laboratory}}
             \vfill}
\hfill
\vbox to 1in {\vfill \FermiSmallfont
              \hbox{#1}
              \hbox{#2}
              \hbox{#3}
              \vfill}
}}\vskip-0.0truein\end{center}}

\def\apj{Astrophys.\ J.\ }
\def\apjl{Astrophys.\ J.\ Lett.\ }
\def\mnras{Mon. Not. R. Astr. Soc.\ }
\def\aap{Astron. Astrophys.\ }

\begin{document}

\title{Bulk QCD thermodynamics and sterile neutrino dark
matter}

\author{Kevork N. Abazajian}
\email{aba@fnal.gov}
\affiliation{NASA/Fermilab Astrophysics Group, Fermi National
Accelerator Laboratory, Box 500, Batavia, Illinois 60510-0500}
\author{George M. Fuller}
\email{gfuller@ucsd.edu}
\affiliation{Department of Physics, University of California, San Diego,
La Jolla, California 92093-0319}
\date{April 17, 2002}

\pacs{95.35.+d,14.60.St,12.38.Mh,25.75.-q \hspace{3.4cm} FERMILAB-Pub-02/067-A}

\begin{abstract}
We point out that the relic densities of singlet (sterile) neutrinos
of interest in viable warm and cold dark matter scenarios, depend on
the characteristics of the QCD transition in the early universe.  In
the most promising of these dark matter scenarios the production of
the singlets occurs at or near the QCD transition.  Since production
of the singlets, their dilution, and the disappearance of weak
scatterers occur simultaneously, we calculate these processes
contemporaneously to obtain accurate predictions of relic sterile
neutrino mass density.  Therefore, a determination of the mass and
superweak mixing of the singlet neutrino through, for example, its
radiative decay, along with a determination of its contribution to the
critical density, can provide insight into the finite-temperature QCD
transition.
\end{abstract}

\FNALppthead{NASA/Fermilab Astrophysics
  Center}{astro-ph/0204293}{FERMILAB-Pub-02/067-A} 

\maketitle

\section{Introduction}
In this paper we show how the temperature and the order (first order
or not) of the QCD transition in the early universe is important in
determining the relic densities of sterile neutrinos in models where
they provide the dark matter.  The QCD transition epoch is the regime
where the quark and gluon degrees of freedom annihilate, chiral
symmetry is broken, and quarks are incorporated into color singlets.
For the low chemical potential (i.e., small baryon number or
high entropy) characterizing the universe, we expect the QCD epoch to
occur at a temperature (energy scale) of order of the pion mass, $T_{\rm
crit} \sim 100\rm\ MeV$.  Coincidentally, this is the temperature
regime where the dark matter singlet neutrinos are produced by
scattering-related processes.

Therefore, an accurate calculation of the density of dark matter
candidate sterile neutrinos requires detailed specification of the
evolution of  the background plasma through the QCD transition.  It
is conceivable that this relation between the QCD transition and
sterile neutrino properties could be exploited to understand the
cosmic QCD epoch.  This approach is not available, for example, with
supersymmetric dark matter candidates which freeze-out of the plasma
at temperatures well above the QCD transition.  Here, we also review
the current best constraints on sterile neutrino dark matter
candidates.

The current cosmological paradigm of a spatially flat
Friedmann-Lemaitre-Robertson-Walker (FLRW) universe with critical
density components shared between cold dark matter [(CDM), $\Omega_m
\sim 0.3$] and a negative-pressure dark energy ($\Omega_X \sim 0.7$),
with nearly scale-invariant adiabatic Gaussian perturbations produced
by inflation leading to structure formation ($\Lambda\rm CDM$), is
supported by a concordance of observational evidence from the cosmic
microwave background, large scale structure, rich galaxy clusters, and
high-redshift Type-Ia supernovae~\cite{cosmo101}.  There is much
optimism in observational cosmology, but a number of unresolved
problems remain.  Among them is the very nature of particle dark
matter and dark energy.  

More astrophysically, there may be a discrepancy between the quantity
of substructures on galactic scales predicted in numerical simulations
using $\Lambda\rm CDM$ initial conditions and that observed in the local
group: the ``dwarf-galaxy problem''~\cite{Moore:1999wf,Ghigna1999}.
Another discrepancy exists between the observed flat and low density
profiles of dwarf galaxy cores~\cite{Dalcanton:2000hn,Bullock2001}
relative to those found in $N$-body simulations.  One solution to the
dwarf-galaxy problem is a modification of the primordial power
spectrum at sub-galaxy scales by thermal damping of perturbations by
warm dark matter (WDM)~\cite{wdm}.
Dwarf-galaxy halos in WDM scenarios form through fragmentation of
larger structures, and the concentrations of their halos may be
reduced~\cite{AvilaReese2001}.  Although these problems may be
signaling the presence of WDM, the apparent dwarf-galaxy problem may
be the result of dispersion of baryons from halos due to late
reionization or supernovae, so that the unobserved dwarf halos are
simply dark. These issues could be resolved through refined
calculations of star formation in small halos, or through observations
of galactic substructure in, for example, gravitational lensing
studies~\cite{dalal}.

Sterile neutrinos are a natural candidate for
WDM~\cite{Dodelson:1994je,Dolgov:2000ew} and
CDM~\cite{Shi:1998km,Abazajian2001}, and can emerge from models with composite
fermions~\cite{Arkani-Hamed:1998pf}, mirror fermions \cite{Berezhiani:1995yi}
or light axinos~\cite{Chun:1999cq}.  Sterile neutrinos with masses of
$1-100\rm\ keV$ and small mixings with active neutrinos are produced in a
nonequilibrium quantum decoherence process associated with the scattering of
active neutrinos: either with the usual simplifying assumption of a small
universal lepton number~\cite{Dodelson:1994je} (of order of the baryon number)
or with the more liberal possibility of a large lepton number in neutrinos
(several orders of magnitude larger than the baryon number)~\cite{Shi:1998km}.
We define the lepton number in a neutrino flavor here as the difference between
neutrino and antineutrino number densities normalized by the photon number
density, $n_\gamma$, at a given epoch: $L_{\alpha} \equiv (n_{\nu_\alpha} -
n_{\bar\nu_\alpha})/n_\gamma$.

Constraints on the lepton numbers which reside in the neutrino seas can be made
by requiring successful production of the light element abundances in
primordial nucleosynthesis while keeping the baryon density compatible with
that inferred from observations of the cosmic microwave
anisotropies~\cite{degenerates}.  Because of the near bi-maximal mixing
paradigm emerging for the neutrino mass matrix, more stringent constraints can
be placed on the neutrino lepton number since asymmetries in the more poorly
constrained mu and tau flavors can be converted to an asymmetry in the more
stringently constrained electron neutrinos through synchronized flavor
transformation~\cite{nudegen}.  However, even the strongest constraint
available, on the electron neutrino lepton number, allows $|L_e|_{1\rm\,MeV}
\lesssim 0.1$ at the epoch of nucleosynthesis.  Because sterile neutrinos are
produced at temperatures much higher than those characterizing big-bang
nucleosynthesis (BBN), and because the lepton number (unlike degeneracy
parameter) is not a comoving invariant, we quote values of the lepton number at
the epoch where $T=100\rm\ GeV$.  Therefore, the most conservative upper bound
on the lepton number at this temperature is weaker by the ratio of the number
of statistical degrees of freedom $g(T)$ in the plasma, $g({100\rm\
GeV})/g({1\rm\ MeV})$, or $|L_e|_{100\rm\,GeV} \lesssim 1$.  All
lepton-number-driven mechanisms we consider are well within this conservative
limit.

Depending on their masses and non-thermal energy distributions, the
sterile neutrinos produced in the early universe can be CDM, WDM, or
hot dark matter (HDM).  The production mechanisms and constraints on
these models were studied in Refs.~\cite{Dolgov:2000ew,Abazajian2001}.
The strongest constraints on the upper bound of the mass of the
sterile neutrino derive from x-ray emission from nearby galaxy
clusters, viz. the Virgo cluster~\cite{Abazajian:2001vt}, $m_s
\lesssim 5\rm\ keV$ when $L\sim 10^{-10}$.  Lower bounds on the mass
of the WDM particle can be made by requiring that its free-streaming
not excessively damp out the small-scale structure, i.e., behave too much
like HDM.  Limits of this type come from the observed structure on
small scales in the high-redshift Lyman-$\alpha$
forest~\cite{Narayanan2000} and by requiring sufficiently early
reionization of the universe seen by the Gunn-Peterson
effect~\cite{Barkana:2001gr}.  A lower bound on the mass can also be
derived from the observed phase-space densities of galaxy
cores~\cite{Dalcanton:2000hn}.  All of these limits require
out-of-equilibrium (``standard'') WDM particles with mass $m \gtrsim
0.75\rm\ keV$, or sterile neutrino WDM in models where $L\sim
10^{-10}$ with mass $m_s \gtrsim 2.6 \rm\ keV$~\cite{Hansen:2001zv}.
It is exciting that the currently unconstrained mass range $2.6 {\rm\
keV} \lesssim m_s\lesssim 5 {\rm\ keV}$ (at $L\sim 10^{-10}$) is that
favored by WDM solutions to small scale structure formation, and, even
more, this mass range for the sterile neutrino dark matter is also
eminently detectable via radiative decay~\cite{Abazajian:2001vt}.

\section{The Quark-Hadron Transition}

The study of the condensed matter physics of QCD remains an active
topic, through analysis of the full color $SU(3)$ QCD Lagrangian with
numerical lattice simulation and analogy with spin systems (for a
recent survey see Ref.~\cite{Satz:2000hm}).  The nature of the QCD
transition at finite temperature and zero chemical potential $\mu_q$
is clear in models containing two massless quarks
($m_{q_u}=m_{q_d}=0$, $m_{q_s}\rightarrow \infty$): the transition is
second order~\cite{Pisarski:1984ms}.  With physically realistic small
but nonzero $m_{q_u}$ and $m_{q_d}$, the second order transition
loses its critical behavior and is a smooth
crossover~\cite{Berdnikov:1999ph}.  In degenerate three-flavor models
($m_{q_u}=m_{q_d}=m_{q_s}=0$), it was inferred analytically that the
transition is first order~\cite{Pisarski:1984ms}, and this has been
supported by lattice studies~\cite{Brown:1990ev}.  Therefore, with
$\mu_q\rightarrow 0$ (as is appropriate for the early universe where
$\mu_q/T\sim 10^{-8}$) there exists some critical mass $m_{q_s}^c$,
below which the transition is first order.  The position of this
critical point in temperature, $\mu_q$, and $m_{q_s}$ is still under
investigation by lattice simulations and may be probed by relativistic
heavy ion collision experiments~\cite{Berdnikov:1999ph}.  Lattice
calculations indicate $m_{q_s}^c$ is well below the inferred physical
strange quark mass~\cite{Brown:1990ev,Karsch:2001nf}, but these
results are not conclusive~\cite{Iwasaki:1996zt}.  Since production of
sterile neutrinos peaks at $T \sim 130 {\rm\ MeV}(m_s/3{\rm\
keV})^{1/3}$, and the QCD transition likely occurs from $T_{\rm crit}
= 100-200\rm\ MeV$, the effects of the transition on production are
unmistakably important.

Convincing theoretical motivation or experimental detection of the
critical end point as yet does not exist, and so whether the QCD
transition in the early universe behaves as condensing water
(first order) or thickening pudding (smooth crossover) remains to be
discovered.  We argue below how the dark matter, if sterile neutrinos,
can also be used to probe the critical behavior of QCD.

We model the evolution of the universe through a first-order
transition as in Refs.~\cite{Alcock:1987tx,Fuller:1988ue}.  The
transition from the high-temperature quark-gluon phase to the hadron
phase includes the change in the statistical degrees of freedom of the
plasma, the release of vacuum energy, and the annihilation of
quark-antiquark pairs.  Supercooling of the plasma can lead to
nucleation of remaining quarks into dense isothermal baryon number
fluctuations~\cite{Witten:1984rs}, but this does not affect the
analysis here.

A first-order transition evolves through a constant-temperature epoch, with a
duration of the order of a Hubble time at that epoch.  During this constant
temperature, mixed phase regime, volume is swapped from the quark-gluon phase
to bubbles of the hadronic phase as the universe expands.  In this scenario the
fraction of the volume in the quark-gluon phase evolves from $f_v = 1$ to $0$,
at which time the transition ends.  The two phases differ in entropy density
and therefore latent heat is released in the transition.  The dynamics of the
scale factor, $a$, through the transition is given
by~\cite{Kajantie:1986hq,Fuller:1988ue}
\begin{equation}
\frac{\dot a}{a} = \chi \left(4 f_v + \frac{3}{x-1}\right)^{1/2},
\label{scalea}
\end{equation}
where $x\equiv g_q/g_h$ is the ratio of the number of relativistic
degrees of freedom in the quark $g_q$ and hadron $g_h$ phases
(approximately valid during the mixed phase epoch), and the
characteristic QCD expansion rate is
\begin{equation}
\chi \equiv \left(\frac{8\pi B}{3 m_{\rm Pl}^2}\right)^{1/2} \approx
  (143{\rm\ \mu s})^{-1} \left(\frac {T_{\rm crit}}{100{\rm\
  MeV}}\right)^2.
\end{equation}
Here the vacuum energy is parametrized by an effective ``bag
constant'' $B$.  Pressure equilibrium between quark and hadron phases
is used to find the relation between the vacuum energy and
the transition's critical temperature, $T_{\rm crit}$.

\begin{figure}
\includegraphics[width=3.375in]{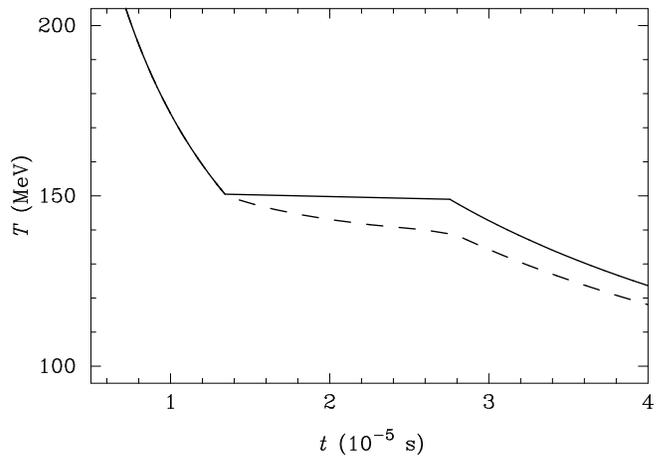}
\caption[]{\label{timetemp} Evolution of temperature for a first-order
  transition (solid) and crossover transition (dashed) for $T_{\rm
  crit} = 150\rm\ MeV$. }
\end{figure}

Using the conservation of comoving entropy it can be shown that the
fraction of the plasma in the quark-gluon phase at a given time, $t$,
from the start of the transition at time, $t_i$, is~\cite{Fuller:1988ue}
\begin{eqnarray}
\label{fv}
f_v &=& \frac{1}{4(x-1)}\Bigg\{ \tan^2\Bigg[ \arctan
    \left[(4x-1)^{1/2}\right]\nonumber \\*
&& + \frac{3 \chi
    (t_i-t)}{2(x-1)^{1/2}}\Bigg] - 3\Bigg\}.
\end{eqnarray}
These relations are used to calculate the cosmological dynamics for
sterile neutrino dark matter production during the first-order
transition.  The evolution of temperature for a first-order and
crossover transition is shown in Fig.~\ref{timetemp}.

\section{Dark Matter Production}

Production of sterile neutrino dark matter occurs through decoherence
of the active neutrino gas via collisions with the background plasma
of weakly interacting particles.  We calculate this nonequilibrium
production of sterile neutrinos as described in
Ref.~\cite{Abazajian2001}, which we summarize here.  This 
production is described by a semiclassical Boltzmann equation that
captures the quantum behavior of the system.  The distribution
function for sterile neutrinos $f_s$ over momenta $p$ evolves as
\begin{align}
&\frac{\partial}{\partial{t}}f_s(p,t) - H\,p\,
\frac{\partial}{\partial{p}}f_s(p,t)
\nonumber\\ 
&\qquad = \Gamma(\nu_\alpha\rightarrow\nu_s;p,t)
\left[f_\alpha(p,t)-f_s(p,t)\right].
\label{boltzmann}
\end{align}
Here $f_\alpha$ is the active neutrino distribution function that is
coupled to the sterile, $\Gamma(\nu_\alpha\rightarrow\nu_s;p,t)$
is the effective rate of production (annihilation) of sterile states
and $H=\dot a/a$. 

It is important to note that the scattering kernels
$\Gamma(\nu_\alpha\rightarrow\nu_s;p,t)$ for the case of sterile
neutrino production are always momentum conserving~\cite{McKellarBell}
so that Eq.~(\ref{boltzmann}) is exact.  We handle the calculation of
this scattering kernel as described in Ref.~\cite{Abazajian2001}.
This includes augmenting the scattering rate as the population of weak
scatterers increases at high temperatures.

The behavior of the QCD transition affects dark matter production
through the varied evolution of the total energy density, $\rho_{\rm
tot}$, of the plasma for different critical behavior.  In turn, this
modifies the evolution of the expansion rate through the Friedmann
equation $H = \sqrt{8\pi/3} m_{\rm Pl}^{-1} \rho_{\rm tot}^{1/2}$.
The dynamics of expansion enters the calculation of dark matter
production in two ways: (1) alteration of the time-temperature
relationship; (2) the modification of redshifting of relativistic
species, including sterile neutrinos.

\begin{figure}
\includegraphics[width=3.375in]{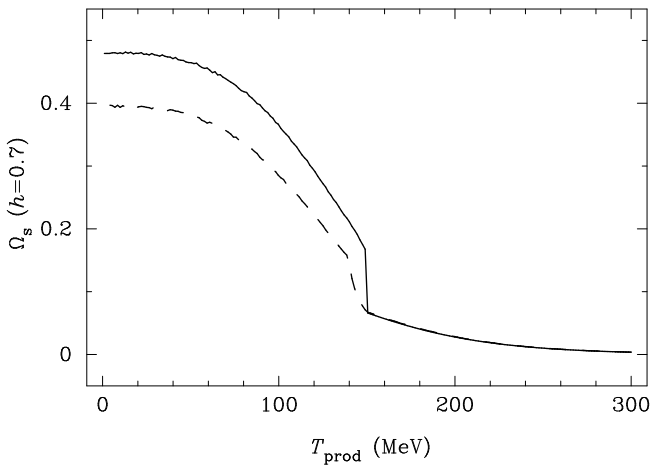}
\caption[]{\label{omegavstemp}The fraction of critical density
$\Omega_s$ produced above a temperature $T_{\rm prod}$.  Here we have
taken $T_{\rm crit} = 150\rm\ MeV$. There is significant enhancement
due to thermal lingering after $T_{\rm crit}$ for a crossover
transition (dashed), and there is a dramatic enhancement at $T_{\rm
crit}$ in the first order case (solid).  The most significant
difference between the two cases stems from the delay of the
radiation-dominated era in a first-order transition.  Here we have
taken $m_s = 3\ \rm keV$, $\sin^2 2\theta=2.6\times 10^{-8}$, and
$L\sim 10^{-10}$. }
\end{figure}

If the universe underwent a crossover QCD transition (i.e., no phase
transition), the effects on sterile neutrino densities stem entirely
from the annihilation of quark-antiquark pairs and the resulting
disappearance of degrees of freedom in the plasma.  The entropy
carried in the annihilating species is transferred to the remaining
constituents of the plasma.  The time-temperature relation in the
early universe with evolving numbers of thermalized species is far
from a new problem.  This is basically the same effect (though more
pronounced) on the time-temperature relation as that caused by the
annihilation of electron-positron pairs when $T\sim 25\rm\ keV$. A
general treatment of the time-temperature relationship including the
effects of the population of sterile neutrinos, is given in the
Appendix of Ref.~\cite{Abazajian2001}, whose approach we have
incorporated.  A crossover transition disturbs production by causing
the universe to ``linger'' for some time at temperatures just below
the critical temperature (see Fig.~\ref{timetemp}).  Since the QCD
transition temperatures can be at or near the peak rate of dark matter
production, the position of the post-critical temperature lingering
may enhance the production.  A representative case of dark matter
production through a crossover transition is shown in
Fig.~\ref{omegavstemp}.

The temperature is constant in a first-order transition, and the duration of
the transition is given by the time required for the quark phase to disappear,
Eq.~(\ref{fv}).  Here, the production with temperature is a step-like function
at the critical temperature as a result of the finite time spent at $T_{\rm
crit}$ (see Fig.~\ref{omegavstemp}).  More significant than the production
during the transition is the delay of the radiation domination epoch by a
first-order transition, so that the time-temperature relation is modified from
the standard case, $t\propto T^{-2}$.  The duration of the first-order
transition can be obtained by Eq.~(\ref{fv}),
\begin{equation}
\Delta t = \frac{2(x-1)^{1/2}}{3\chi } \left\{
  \arctan\left[4(x-1)^{1/2}\right] - \arctan 3^{1/2}\right\}\,,
\end{equation}
which for typical parameters is approximately $2\times 10^{-5}\rm\ s$.
A first-order transition universe will then spend slightly more time
reaching a given temperature.  This can be seen in
Fig.~\ref{timetemp}.  This delay in the temperature decrease with
expansion is a major factor in enhancing dark matter production in a
first-order transition universe.  That is, the partial equilibration
of the sterile neutrinos has more time to operate in the first-order
case.

In this way, the different dependence of the scale factor with time gives rise
to different redshift histories in different transition scenarios. The
expansion also modifies the redshifting of sterile neutrino momenta ($p\propto
a^{-1}$). In a radiation-dominated environment, and with a crossover
transition, $p \propto t^{-1/2}$.  However, during a first-order phase
transition the momentum will scale inversely with the scale factor as found by
integrating Eq.~(\ref{scalea}).  Handling the redshifting explicitly through
these scalings as opposed to using the second term on the left-hand side of
Eq.\ (\ref{boltzmann}) greatly accelerates numerical computation by eliminating
the calculation of the momentum derivative of the time-evolving sterile
neutrino distribution.  In this way, the different dependence of the scale
factor with time gives rise to different redshift histories in different
transition scenarios.

\begin{figure}
\includegraphics[width=3.375in]{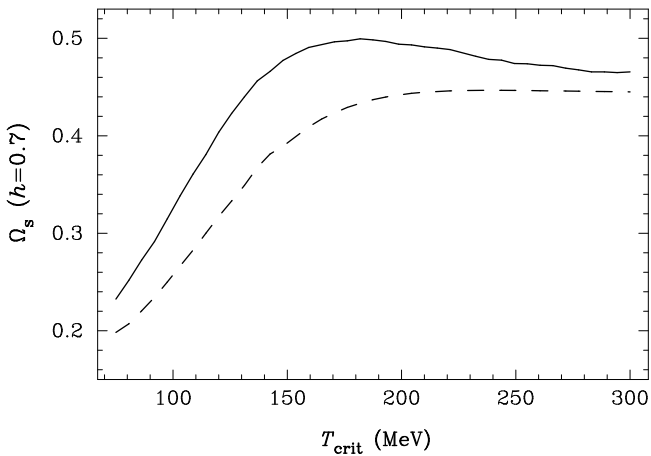}
\caption[]{\label{omegavstc} Fraction of critical density contributed by relic
  sterile neutrinos for a varying critical temperature for the QCD transition.
  Solid line is for a simple crossover QCD transition; dashed line is for a
  first-order transition. Here we have specified $m_s = 3\ \rm keV$, $\sin^2
  2\theta=2.6\times 10^{-8}$, and $L\sim 10^{-10}$. }
\end{figure}

The final effect of the transition we consider is sterile neutrino dilution by
the heating of the thermally coupled species in the plasma relative to the
uncoupled dark matter by entropy transfer (latent heat release).  Exclusive of
all other effects in a crossover transition, this produces a heating of the
coupled species relative to the dark matter.  With the assumption that the
crossover transition is rapid relative to the expansion time, the dilution of
the dark matter relative to coupled species is given by comoving entropy
conservation as $x^{-1} =g_h/g_q \sim 1/3$.  A crossover transition does not
necessarily take place in less than the expansion time scale, as can be seen
for the cases we consider in Fig.~\ref{timetemp}.  Therefore, the dilution
should be handled by a Boltzmann formulation as is automatically done in our
treatment of Eq.~(\ref{boltzmann}) and the time-temperature relation.  For a
crossover transition, dilution is the major factor in determining the different
critical density contributions shown in Fig.~\ref{omegavstc}.  In a first-order
transition, dilution effects are handled in the evolution of the scale factor
(\ref{scalea}) and time-temperature relation.

\begin{figure}
\includegraphics[width=3.375in]{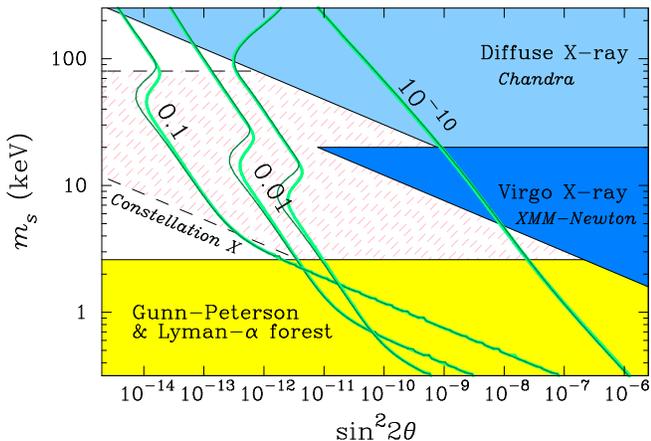}
\caption[]{\label{param} Shown is the parameter space available for
sterile neutrino dark matter, with varying lepton-number cosmologies.
The contours (numbered by their initial lepton number) are positions
in the mass and mixing angle space where sterile neutrinos produce
critical densities of $\Omega_s h^2 = 0.15$.  The thin (thick) lines
are for first-order (crossover) QCD transitions ($T_{\rm crit} =
150\rm\ MeV$).  Also shown are the excluded regions (shaded gray) from
small scale structures---the Gunn-Peterson bound and Lyman-$\alpha$
forest---and halo phase space densities, the resolution of the diffuse
x-ray background by {\it Chandra}, and observations of the Virgo
cluster by {\it XMM-Newton}.  The dashed region is that which may be
probed by the proposed {\it Constellation-X} mission~\cite{Abazajian:2001vt}.}
\end{figure}

We have included all of the above effects in a numerical solution of
the production equations (\ref{boltzmann}).  Contours of predicted
sterile neutrino critical densities of $\Omega_s h^2 = 0.15$ for three
representative lepton number cosmologies and both first-order and
crossover QCD transitions are shown in Fig.~\ref{param}.  The
shape of the large lepton number cosmology contours can be understood
from the resonance condition for a neutrino of momentum $p$:
\begin{eqnarray}
\frac{p}{T}\Big|_{\rm res} &=& \frac{\delta m^2 \cos
  2\theta}{8\sqrt{2} \zeta (3) G_F L_\alpha T^4 /\pi^2} \cr
&\approx& 0.3 \left(\frac{m_s}{1\rm\ keV}\right)^2
  \left(\frac{0.01}{L_\alpha}\right) \left(\frac{150\rm\
  MeV}{T}\right)^4\,.
\end{eqnarray}
The largest disparity between first-order and crossover transitions
occurs when the resonance position is in a populated portion of the
neutrino momentum distribution (of order $p/T\sim 1$) at $T_{\rm
crit}$.  For an $L=0.01$ cosmology, the resonance is in the peak of
the momentum distribution at $T_{\rm crit}$ for masses $m_s \approx
13\rm\ keV$, the pronounced effects of the nature of the transition
for this case can be seen in Fig.~\ref{param}.  Furthermore, for this
lepton number, the resonance position falls at very low momenta for
$m_s < 1\rm\ keV$ during the peak production.  Therefore, production
is thermally suppressed, leading to the break seen at $m_s \sim 1\rm\
keV$ for $L=0.01$ cosmologies in Fig.~\ref{param}.

An effect present above the quark-hadron transition is that of quark screening,
or many-body neutrino-quark scattering.  The mean distance between quarks above
the quark-hadron transition at $T\sim 150 \rm \ MeV$ is approximately $l\sim[3
\zeta(3) g_q/\pi^2]^{-1/3} T^{-1} \sim 1\rm\ fm$, and the de Broglie wavelength
of a typical neutrino is $\lambdabar\sim \hbar/p \sim 0.4 \rm\ fm$.  Therefore,
neutrinos below the average momentum will have $\lambdabar > l$, and two body
single-particle scattering is no longer a complete description.  Many-body
coherent scattering will tend to cancel the weak charge of quarks and
antiquarks, and reduce the effectiveness of decohering interactions of active
neutrinos.  The effects of many-body scattering are beyond the scope of this
work. Though it can be significant in altering scattering rates above the QCD
transition, the majority of sterile neutrino dark matter is produced below the
QCD transition (as seen in Fig.~\ref{omegavstemp}) and the resulting effects of
quark screening will be relatively small.

\section{Probing the QCD transition with Dark Matter}

In order to predict the sterile neutrino dark matter density one must
specify the singlet neutrino rest mass $m_s$, vacuum mixing with
active neutrinos $\sin^2 2\theta$, the initial lepton number $L$, and
the order of the QCD transition and its critical temperature $T_{\rm
crit}$.  Since the dark matter density is being determined rather
precisely~\cite{cmbdm}, the sterile neutrino production mechanism will
constrain the relationship between the sterile neutrino properties and
parameters describing the background plasma.

It was shown in Ref.~\cite{Abazajian:2001vt} that the radiative decay
of sterile neutrino dark matter $\nu_s\rightarrow\nu_\alpha\gamma$ may
be detected with high-sensitivity spectrographs aboard the current
{\it Chandra} or {\it XMM-Newton} x-ray telescopes, or with the
proposed large-surface-area and precision spectrograph of the {\it
Constellation X} mission.  As noted earlier, the lack of a significant
line in the observation of the Virgo cluster constrains the largest
possible mass of a sterile neutrino dark matter candidate produced in
the zero-lepton-number case ($m_s \lesssim 5\rm\ keV$).  This
constraint comes from the fact that the Virgo cluster is one of the
better candidate objects~\cite{Abazajian:2001vt} and a
high-sensitivity spectrum is available~\cite{boehringer}.  Observations
of this and other clusters and field galaxies with current x-ray
observatories can either detect or exclude sterile neutrino dark
matter as a dark matter candidate in zero-lepton-number universes.

If sterile neutrinos are the dark matter, detection of radiative decay
by a sufficiently sensitive observatory would readily identify $m_s$
since the radiative decay produces monoenergetic photons with energy
$E_\gamma = m_s/2$.  Detection of these photons in the x-ray is most
likely in long-duration exposures of nearby high surface dark matter
mass density objects.  The x-ray flux coming from the 
radiative decay of a sterile neutrino is
\begin{eqnarray}
\label{flux}
F &\approx& 5.1\times 10^{-18} {\rm\ erg\ cm^{-2}\
  s^{-1}}\left(\frac{D}{\rm Mpc}\right)^{-2} \left(\frac{M_{\rm
  DM}}{10^{11} M_\odot}\right)\nonumber \\
&&\quad \times\left(\frac{\sin^2
  2\theta}{10^{-10}}\right) \left(\frac{m_s}{1\rm\ keV}\right)^5\,.
\end{eqnarray}
Note that the radiative decay rate of a heavy sterile neutrino is greatly
enhanced due to the lack of the Glashow-Iliopoulos-Maiani suppression present
in the decay of active neutrinos of the same mass and
mixing~\cite{palwolfenstein}.

Since the position of the x-ray line will specify $m_s$, the strength
of the line determines $\sin^2 2\theta$, and therefore the position in
Fig.~\ref{param}.  Of the remaining parameters ($L_\alpha$,
$T_{crit}$ and the order of the transition), only a small range will
successfully produce the observed critical density and detected
radiative decay flux.  However, there remains a degeneracy between
$L_\alpha$ and the order of the transition since both a first-order
transition and a slightly larger $L_\alpha$ can enhance production at
smaller mixing angles (see Fig.~\ref{param}).  However, if one can
independently determine either the order of the transition or
$L_\alpha$, then the observed flux [Eq.~(\ref{flux})] will
determine the remaining degree of freedom.

\section{Conclusions}

We have presented the results of the most detailed analysis of sterile
neutrino dark matter production yet performed.  In doing so, we have
explored the implications of an unexpected coincidence: the peak
production epoch for sterile neutrino dark matter can be the same
as the QCD epoch where the quark and gluon degrees of freedom
disappear and give way to color singlets.  The effects of the QCD
transition on sterile neutrino relic densities are bracketed by 
simple crossovers (no phase transition) and first-order phase
transitions for the reordering of the strongly interacting degrees of
freedom from quarks and gluons to hadrons.

We find that the varying alteration in the expansion dynamics in these
different models of the QCD transition at high temperature and low chemical
potential can give rise to an appreciable range of closure contributions
associated with a given singlet sterile neutrino mass, vacuum mixing with
active species, and primordial lepton number(s).  In particular, first-order
phase transition scenarios for the cosmic QCD epoch produce a constant
temperature interval with a duration of the order of the Hubble time associated
with phase coexistence.  This can result in a significant enhancement of
sterile neutrino relic densities relative to simple crossover transitions.

Ultimately, we have pointed out here an unexpected connection between
strong-interaction scale physics in the early universe on the one hand
and low energy scale dark matter relic physics at the present epoch on
the other.  Whether this connection could someday be exploited to
produce insights into the cosmic QCD epoch would, of course, depend on
the identification of a significant component of the dark matter as being
carried by sterile neutrinos. 

An accurate calculation of the sterile neutrino production entails following
the evolution of the active neutrino scatterer number densities, sterile
neutrino production, and entropy transfer-induced dilution for a range of
possible behaviors for the strongly interacting sector.  Including these
effects, we have presented the most sophisticated calculation of the
predictions of sterile neutrino relic densities for a range of possible QCD
transition scenarios and lepton number cosmologies.

\acknowledgments

We would like to thank John Beacom, Nicole Bell, Steen Hansen, Rocky Kolb, and
Krishna Rajagopal for useful discussions, and, in addition, we thank Rocky Kolb
for providing a rapid numerical calculator of the evolution of the number of
statistical degrees of freedom in the early universe.  This research was
supported in part by the DOE and NASA grant NAG 5-10842 at Fermilab and by NSF
Grant PHY00-99499 at UCSD.

\end{document}